\documentclass[10pt,twocolumn,showpacs,eqsecnum,superscriptaddress]{revtex4}

\usepackage[intlimits]{amsmath}
\usepackage{amsfonts, amssymb, amsthm}

\usepackage{graphicx}
\usepackage{subfigure}
\usepackage{mathptmx}

\newcommand{\crit}{_{\text{c}}}

\newcommand{\tc}{T_c}

\newcommand{\secref}[1]{Sec.~\ref{#1}}

\newcommand*{\bfigref}[1]{Figure~\ref{#1}}  

\newcommand{\mean}[1]{\langle #1 \rangle}
\newcommand{\msd}{\mean{\Delta x^2}}

\newcommand{\defeq}{:=}

\newcommand{\defn}[1]{\emph{#1}}

\newcommand{\e}{\mathrm{e}}
\newcommand{\rd}{\mathrm{d}\,}

\newcommand{\modulus}[1]{\left| #1 \right|}

\newcommand{\figref}[1]{Fig.~\ref{#1}}
\newcommand{\subfig}[2][1]{\subfigure[]{\includegraphics[scale=#1]{#2}}}

\newcommand{\scalefig}[2]{\includegraphics[scale=#1]{#2}}

\newcommand{\Dx}{\Delta x}

\newcommand{\Warwick}{Mathematics Institute, University of Warwick,
Coventry, CV4 7AL, U.K.}  
\newcommand{\Cuernavaca}{Centro de Ciencias
F\'{\i}sicas, UNAM, Apartado postal 48-3, C.P.\ 62551, Cuernavaca,
Morelos, Mexico}

\begin{document}

\title{Occurrence of normal and anomalous diffusion in
polygonal billiard channels}
\author{David P.\ \surname{Sanders}}
\email{dsanders@fis.unam.mx}
\affiliation{\Cuernavaca}
\affiliation{\Warwick}

\author{Hern\'an \surname{Larralde}}
\affiliation{\Cuernavaca}

\begin{abstract}

From extensive numerical simulations, we find that periodic polygonal
billiard channels with angles which are irrational multiples of $\pi$
generically exhibit normal diffusion (linear growth of the mean
squared displacement) when they have a finite horizon, i.e.\ when no
particle can travel arbitrarily far without colliding.  For the
infinite horizon case we present numerical tests showing that the
mean squared displacement instead grows asymptotically as $t \log t$.
When the unit cell contains accessible parallel scatterers, however,
we always find anomalous super-diffusion, i.e.\ power-law growth with
an exponent larger than $1$. This behavior cannot be accounted for
quantitatively by a simple continuous-time random walk model. Instead,
we argue that anomalous diffusion correlates with the existence of
families of propagating periodic orbits. Finally we show that when a
configuration with parallel scatterers is approached there is a
crossover from normal to anomalous diffusion, with the diffusion
coefficient exhibiting a power-law divergence.

\end{abstract}
\date{\today}


\maketitle


\section{Introduction}

\label{sec:intro}

Billiard models, in which point particles in free motion undergo
elastic collisions with fixed obstacles, have simple microscopic
dynamics but strong `statistical' behavior, such as diffusion, at
the macroscopic level. The statistical properties of scattering
billiards such as the Lorentz gas, which have smooth, convex
scatterers, are well-understood: see e.g.\ \cite{Szasz} for a
collection of reviews.  Recently, however, models with
\emph{polygonal} scatterers have attracted attention, since they are
only weakly chaotic, with for example all Lyapunov exponents being
$0$, but they still exhibit surprising statistical properties in
numerical experiments \cite{DettCoh1,AlonsoPolyg,LiHeatLinearMixing}.

The dynamics in polygonal billiards depends strongly on the angles of
the billiard table. Some rigorous results are available for angles
which are all \emph{rational}, i.e.\ rational multiples of $\pi$: see
\cite{GutkinRegChaoticDynReview} and references therein. In
particular, an initial condition with a given angle cannot explore the
whole phase space, since the possible angles obtained in the resulting
trajectory are restricted, so that the model does not have good
ergodic properties.  Previous numerical work has shown that if one
angle is rational and the others irrational, then the results are
sensitive to the value of the rational angle \cite{AlonsoPolyg,
LiHeatLinearMixing}.

In this work we restrict our attention to the case in which all angles
are \emph{irrational} multiples of $\pi$.  There are few rigorous
results available in this case, but there
is numerical evidence that the dynamics is mixing in irrational triangles 
\cite{CasatiProsenMixingTriBilliard}.

We have performed extensive simulations of the transport properties of
quasi-one-dimensional polygonal channel billiards; here we report detailed
results for two classes of model. In both we find that transport
\emph{generically} corresponds to what is usually referred to as
normal diffusion, i.e.\ linear growth of the mean squared
displacement, when the particles cannot travel arbitrarily long
distances without collisions (finite horizon).  By `generic' we mean
that the set of models which do not exhibit normal diffusion is
`small', for example having measure zero in the space of geometrical
parameters. Similarly, for the infinite horizon case we find that
transport is generically {\it marginally} super-diffusive, i.e.\ there
is a logarithmic correction to the linear growth law of the mean
squared displacement.

However, we also find exceptions to the generic behaviors described
above.  These exceptions occur when there are \emph{parallel
scatterers} in the unit cell which are accessible from one another
(see \secref{sec:anomalous-diffusion} for the definition).  In such
cases we always find super-diffusion, i.e.\ the mean squared
displacement grows like a power law with exponent greater than
$1$. Further, we demonstrate that as a parallel configuration is
approached, the transition from normal to anomalous diffusion occurs
through a power-law divergence of the diffusion coefficient.  We also
show that the exponents we obtain are not consistent with a simple
continuous-time random walk model of anomalous diffusion.

Our results are consistent with previous observations of irrational
models exhibiting normal
\cite{AlonsoPolyg,AlonsoPolyII,LiHeatLinearMixing} and anomalous
\cite{ZaslavskyEdelmanMaxwell, ZaslavskyReview, LiFiniteThermCond,
ProsenAnomDiffn} diffusion. Also, very recently, numerical results
corresponding to one of the models studied in this work were reported
in \cite{RondoniZigzagPreprint}, where it was also observed that
parallel scatterers result in anomalous diffusion; the results we
obtain complement and extend those of that reference.

Energy transport in polygonal billiards placed between two heat
reservoirs at different temperatures has also been a focus of
attention \cite{AlonsoPolyg,AlonsoPolyII,LiHeatLinearMixing}.  This
process is closer to ``color diffusion'' than to heat conduction
due to the absence of particle interactions and, hence, of local
thermodynamic equilibrium \cite{DharAbsenceLocalThermEq}.  In these
systems the properties of the energy transport process are known to be
closely related to those of the particle diffusion process
\cite{LiNormAndAnomDiffn,DenisovDynamHeatChannels}, and for this
reason we only study the latter.

\subsection{Models}

\begin{figure}
\subfig[0.85]{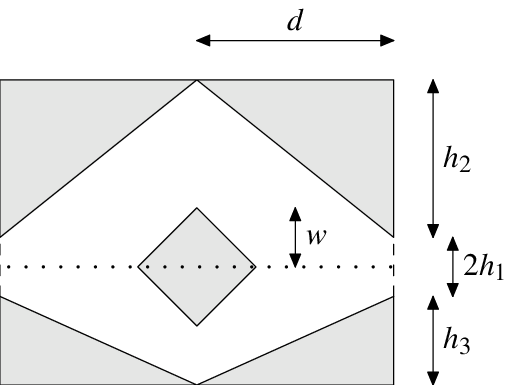}\hfill
\subfig[0.85]{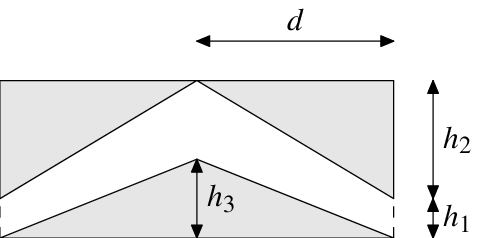} 
\caption{\label{fig:poly-billiard-mine} Unit cells of (a) the
polygonal Lorentz model, and (b) the zigzag model.
\label{fig:zigzag-channel}
}
\end{figure}

The unit cells of our models are shown in
\figref{fig:poly-billiard-mine}; the complete channel consists of an infinite horizontal periodic
repetition of such cells.  The \emph{polygonal Lorentz model}
is a polygonal channel version of the Lorentz gas channel studied in
\cite{Sanders2005}, consisting of a quasi-1D channel with an extra
square scatterer at the center of each unit cell. By \emph{quasi-1D}
we mean that particles are confined to a strip infinitely extended in
the $x$-direction but of bounded height in the $y$-direction.

The polygonal Lorentz model can be simplified by eliminating the
central scatterer and flipping the bottom line of scatterers to point
upwards, resulting in the quasi-1D \emph{zigzag model}. Both models
are designed to permit a \emph{finite horizon} by blocking all
infinite corridors in the structure, so that there is an upper bound
on the distance a particle can travel without colliding with a
scatterer; in scattering billiards this condition was shown to be
necessary for normal diffusion to occur \cite{BSC, Bleher}.  The
zigzag model has also been studied in \cite{RondoniZigzagPreprint}.

We remark that the zigzag model with parallel sides can be reduced to
a parallelogram with irrational angles.  This simple billiard geometry
seems to have been neglected previously, although it is close to a
model considered in \cite{KaplanWeakQuantErg} which was reported to
exhibit a very slow exploration of phase space.  We have found similar
behavior in the parallel zigzag model (see also
\cite{RondoniZigzagPreprint}), but it appears to be a peculiarity of
this model which is not necessarily shared by other polygonal
billiards exhibiting anomalous diffusion.

\subsection{Simulations}

In our simulations we distribute $N=10^7$ initial conditions (unless
otherwise stated) uniformly with respect to Liouville measure, i.e.\
with positions distributed uniformly in the available space in one
unit cell, and velocities with unit speed and uniformly distributed
angles. We consider only irrational angles for the billiard walls in
each model, although we vary explicitly scatterer heights rather than
angles, and to fix the length scale we take $d=1$ (see
\figref{fig:poly-billiard-mine}) .

We study statistical properties of the particle positions $x(t)$ as a
function of time $t$, denoting averages at time $t$ over all initial
conditions by $\mean{\cdot}_t$.  Of particular interest is the mean
squared displacement $\msd_t$, where $\Delta x_t \defeq x(t) - x(0)$,
which is frequently used to characterize the transport properties of
such systems.

\section{Normal and marginally anomalous diffusion}
\label{sec:normal-diffn}

In this paper, by \emph{normal diffusion} we mean asymptotic linear
growth of the mean squared displacement $\msd_t$ as a function of time
$t$, which we denote by $\msd_t \sim t$ as $t \to \infty$.  This is
characteristic of systems commonly thought of as `diffusive', for
example random walkers which take uncorrelated steps of finite mean
squared length.

Strongly chaotic dynamical systems, such as the Lorentz gas, are known
to exhibit normal diffusive behavior \cite{Szasz}; this arises due to
a fast decay of correlations, even between neighboring trajectories,
so that after a short correlation time the dynamics looks `random'.
It is rather surprising, then, that normal diffusion also seems to
appear in non-chaotic polygonal billiards
\cite{AlonsoPolyg,AlonsoPolyII,LiHeatLinearMixing} in which
correlations do not decay exponentially. In these systems, the only
source of `randomization' arises due to divergence of neighboring
trajectories colliding on different sides of a corner of the billiard:
see e.g.\ \cite{vanBeijerenWeaklyChaoticEntropies} for a recent
attempt to characterise this randomization effect.

In the following, we present sensitive tests which provide strong
confirmation that, apart from exceptional cases, the decorrelation
induced by the randomization mechanism mentioned above is indeed
sufficient to give rise to normal diffusion in polygonal billiards.
In particular our tests distinguish the logarithmic correction
(marginally anomalous diffusion) which arises generically in
these systems when they have infinite horizon, as it does in fully
chaotic billiards. 

We will see that exceptions to these behaviors occur when the
billiards present accesible parallel walls. In such cases we argue
that long time correlations persist, related to the existence of
families of travelling periodic orbits, which always give rise to
anomalous diffusion.

\subsection{Growth of second moment}
 
\begin{figure}
\scalefig{0.9}{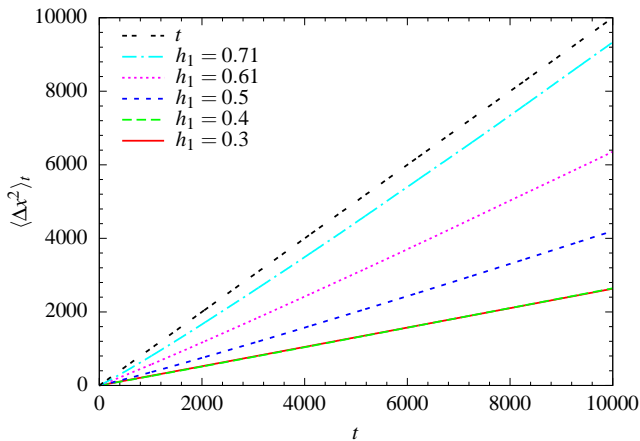} 
\caption{\label{fig:msd} (Color online) $\msd_t$ as a function of $t$
in the zigzag model with $h_2 = 0.77$ and $h_3 = 0.45$.  There is a
finite horizon only when $h_1 \le h_3=0.45$; curvature is barely
visible in the infinite horizon cases.  Statistical errors are smaller
than the line width.  }
\end{figure}

The growth of the mean squared displacement $\msd_t$ as a function of
time $t$ is the most commonly used indicator of transport properties.
In \figref{fig:msd} we plot $\msd_t$ for several non-parallel zigzag
models, varying $h_1$. For values of $h_1$ below $h_3 = 0.45$ there is
a finite horizon and we find that a linear regime is rapidly attained,
suggesting normal diffusion.  For values of $h_1$ above this value,
there is an infinite horizon; in this case, the mean squared
displacement is still almost linear, although a small amount of
curvature is visible.

\begin{figure}
\scalefig{0.9}{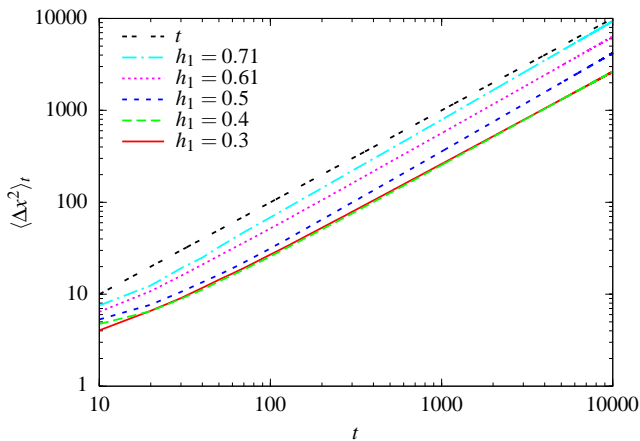} 
\caption{\label{fig:log-log-msd} (Color online) $\msd_t$ as a function
of $t$ on a double logarithmic plot in the zigzag model with the same parameters as in \figref{fig:msd}.}
\end{figure}

Normal diffusion corresponds to an asymptotic slope $1$ of $\msd_t$ on
a double logarithmic plot.  \bfigref{fig:log-log-msd} shows such a
plot for the same data as in \figref{fig:msd}, and we see that indeed
the slope is $1$ for finite horizon and close to $1$ for infinite
horizon (e.g.\ slope $1.06$ for $h=0.71$).  For models with irrational
angles and no accessible parallel sides (see
\secref{sec:anomalous-diffusion}), we always find a growth exponent
close to $1$, so that normal diffusion is generic in polygonal
billiard channels.

The above test does not, however, take account of logarithmic
corrections such as those found in the Lorentz gas with infinite
horizon, where $\msd_t \sim t \ln t$ \cite{Bleher}; such a correction
will appear on the above plot as a small change in the exponent.  A
heuristic argument of \cite{FriedmanMartin84} leads us to expect that
in \emph{any} channel with an infinite horizon, including polygonal
ones, we will have (at least) this kind of `marginal' anomalous
diffusion.  We now present numerical tests which distinguish the
logarithmic correction, showing that the diffusion is normal in the
finite horizon case and marginally anomalous in the infinite horizon
case.

\begin{figure}
\scalefig{0.9}{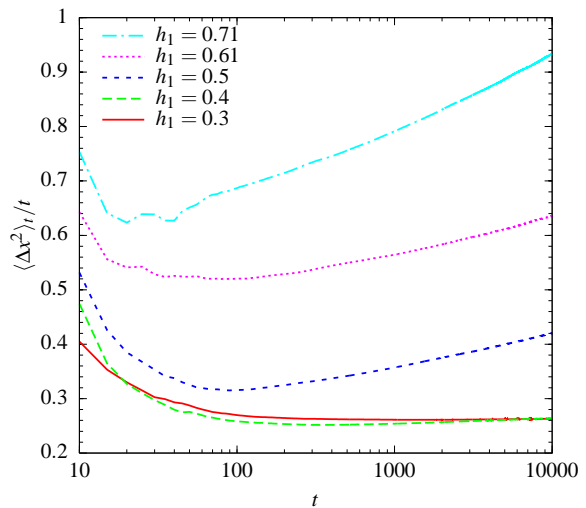} 
\caption{\label{fig:log-msd} (Color online) 
$\msd_t / t$ as a function of $\log
t$ in the zigzag model with the same parameters as in \figref{fig:msd}.}
\end{figure}

In marginal anomalous diffusion we expect that $\msd_t \sim a t \ln t +
b t + c$ for some constants $a$, $b$ and $c$ \cite{GarrGall}, so that
\begin{equation} \label{eq:log-term}
\frac{\msd_t}{t} \sim a z + b + c \e^{-z},
\end{equation}
where $z \defeq \ln t$.  Asymptotic linear growth of $\msd_t / t$ as a
function of $\ln t$ thus indicates marginal anomalous diffusion,
whilst asymptotic flatness corresponds to normal diffusion.  A similar
method was used in \cite{GeiselThomaeAnomDiffusion1984} for a discrete
time system, and we have checked that it also correctly finds marginal
anomalous diffusion in the infinite-horizon Lorentz gas with
continous-time dynamics.  \bfigref{fig:log-msd} plots $\msd_t / t$,
confirming the normal/marginal distinction in finite and infinite
horizon regimes.

We can also fit $\msd_t$ to the expected form $a t \ln t + b t + c$
with parameters $a$, $b$ and $c$, for example using a nonlinear least-squares
method.  For $h_1 = 0.71$ we find $a = 0.051$; this agrees very well
with the asymptotic slope $0.049$ in \figref{fig:log-msd}, as it
should do by equation \eqref{eq:log-term}.  For $h_1 = 0.3$ we instead
obtain $a = -0.0003$, i.e.\ essentially zero (note that it cannot be
negative), confirming normal diffusion.

\subsection{Integrated velocity autocorrelation function}

For a system with normal diffusion we define the diffusion coefficient
$D$ as the asymptotic growth rate $\lim_{t\to\infty} \frac{1}{2t}
\msd_t$ of the mean squared displacement. The Green--Kubo formula $D =
\lim_{t\to \infty} \int_0^t C(\tau) \rd \tau$ \cite{Reichl} relates
$D$ to the velocity autocorrelation function $C(t) = \mean{v_t \,
v_0}$: $D$ exists, and the diffusion is normal, only if $C(t)$ decays
faster than $1/t$ as $t \to \infty$.  $C(t)$ was studied in
\cite{AlonsoPolyII} for a different channel geometry, but this
function is very noisy and it is not possible to determine its decay
rate.

Instead, following \cite{LoweMasters} we consider the integrated
velocity autocorrelation function
\begin{equation}
R(t) \defeq \int_0^t C(\tau) \rd \tau = \mean{v_0 \, \Delta x_t},
\end{equation}
where $\Delta x_t \defeq x(t) - x(0)$ is the particle displacement at
time $t$.  This function is much smoother than $C(t)$, and satisfies
$R(t) \to D$ as $t \to \infty$ if the diffusion coefficient exists,
whilst $R(t) \sim \ln t$ if $C(t)$ decays as $t^{-1}$.

\begin{figure}
\scalefig{0.9}{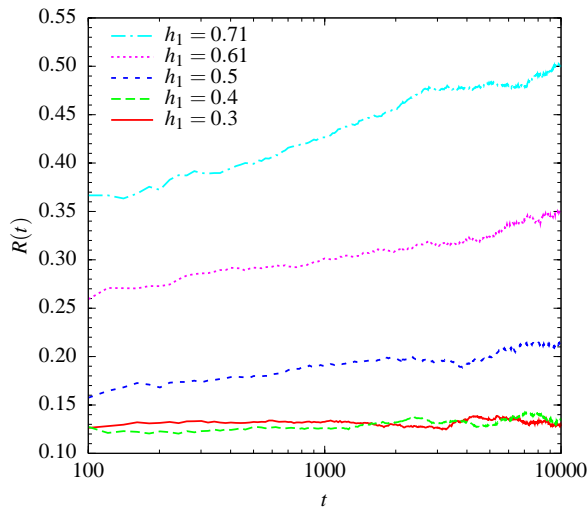} 
\caption{\label{fig:normal-intvaf} (Color online)
Integrated velocity autocorrelation
function $R(t)$ as a function of $\log t$ in the (non-parallel) zigzag
model with the same parameters as in \figref{fig:msd} ($h_2 = 0.77$ and $h_3 = 0.45$).
}
\end{figure}

\bfigref{fig:normal-intvaf} plots $R(t)$ as a function of $\ln t$ for
several finite and infinite horizon non-parallel zigzag models.  The
flatness of $R(t)$ in the finite horizon models indicates that the
limit exists, and hence that the diffusion is normal, and contrasts
with the asymptotic linear growth in the infinite horizon models,
providing further evidence of the marginally anomalous behavior in the
latter case.

\subsection{Higher-order moments}

We define  the growth exponent $\gamma_q$ of the $q$th moment
$\mean{\modulus{x}^q}_t$ by 
\begin{equation}
  \gamma_q \defeq \lim_{t\to\infty} \frac{\log \mean{\modulus{x}^q}_t}{\log
  t},
\end{equation}
which ignores corrections to power-law growth and is a convex function
of $q$; further, $\gamma_q \le q$ for all $q$, since the particles have
finite velocity \cite{ArmsteadOtt}.  The exponents $\gamma_q$, and in
particular the diffusion exponent $\alpha=\gamma_2$, are measured
using a fit to the long-time region of a double logarithmic plot of
the relevant moment, assuming that the asymptotic regime has been
reached.

We find that the behavior of moments of high order ($q \ge 4$) is
dominated by the extreme particles, i.e.\ by those which have the
largest value of $\modulus{x(t)}$, giving data which is not
reproducible between different runs.  For this reason we eliminate the
$5$ most extreme particles from the average before calculating the
growth rate, resulting in data which is now reproducible.  A similar
procedure was used in \cite{ZaslavskyEdelmenWeakMixingChaos2001}.

\begin{figure}
\scalefig{0.9}{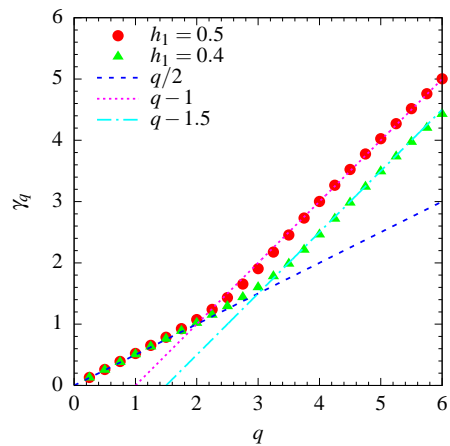} 
\caption{\label{fig:normal-moments} (Color online)
Growth rate $\gamma_q$ of $q$th
moment $\mean{|x|^q}_t$ for finite horizon ($h_1 = 0.4$) and infinite
horizon ($h_1 = 0.5$) in the zigzag model with $h_2 = 0.77$ and $h_3 =
0.45$ for $10^6$ initial conditions.  }
\end{figure}

\bfigref{fig:normal-moments} shows the exponents $\gamma_q$ for the
zigzag model with finite and infinite horizon.  The low-order
exponents satisfy $\gamma_q = q/2$, and in both cases there is a
crossover to a second linear regime with slope $1$, which occurs at
approximately $q = 3$ in the finite horizon case and $q=2$ for
infinite horizon; the latter agrees with the result found in
\cite{ArmsteadOtt} for the infinite horizon Lorentz gas using a method
which also applies here.  Higher-order moments thus provide another
method to distinguish between normal and marginal anomalous diffusion
in polygonal billiards.

The observed qualitative change in behavior of the moments corresponds to a
change in relative importance between diffusive and ballistic effects 
 \cite{ArmsteadOtt, CastiglioneStrongAnomDiffn,
FerrariStrongSelfSimDiffn}, and explains the previous observation that
the Burnett coefficient, defined as the growth rate of the fourth
cumulant $\mean{x^4}_t - 3 \mean{x^2}_t^2$, diverges as a function of
time in polygonal billiards \cite{DettCoh1, AlonsoPolyg}. 

The combination of the above tests provides strong evidence of the
distinction between normal diffusion in finite horizon models and
marginally anomalous diffusion in infinite horizon models.

\section{Anomalous diffusion}
\label{sec:anomalous-diffusion}

\begin{figure}
\scalefig{0.9}{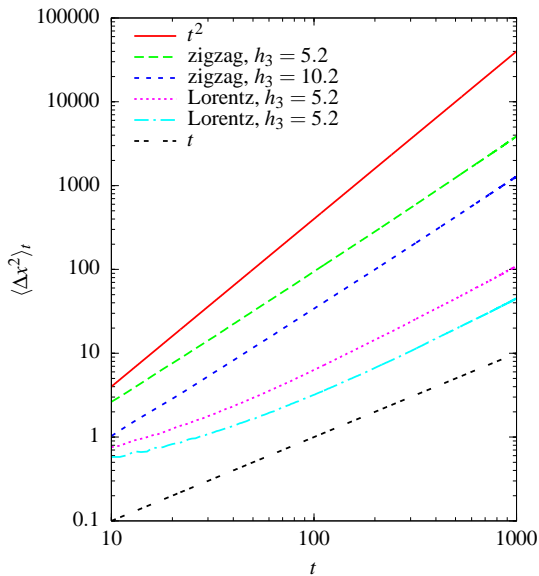} 
\caption{\label{fig:anom-growth} (Color online) Mean squared
displacement $\msd_t$ for parallel models versus $t$ on a double
logarithmic plot.  The upper and lower lines are proportional to $t$
and $t^2$ for comparison. The remaining curves show, from top to
bottom, the parallel zigzag model with $h_1=0.1$ and $h_3=5.2, 10.2$
(both shifted vertically for clarity), and the parallel polygonal
Lorentz model with $h_1=0.1$, $w=0.15$ and $h_3 = 5.2, 10.2$.
Statistical errors are slightly greater than the width of the lines.}
\end{figure}

Whilst we generically find normal diffusion, for certain exceptional
geometrical configurations this does not occur.  The main exception is
when there are \emph{accessible parallel scatterers} in the unit cell
of a model. In this case we always find \emph{anomalous
(super-)diffusion}, i.e.\ $\msd_t \sim t^\alpha$ with $\alpha > 1$:
see \figref{fig:anom-growth}.  Since the particles have finite speed,
we always have $\alpha \leq 2$, with the value $2$ corresponding to
ballistic motion.  Sub-diffusion ($\alpha < 1$) was reported in
\cite{AlonsoPolyg} in a model with one rational angle, but we are not
aware of any model with irrational angles exhibiting sub-diffusion.

We say that a model has `accessible parallel scatterers' when there
are trajectories which can propagate arbitrarily far by colliding only
with scatterers which are parallel to some other scattering element in
the unit cell.

As an example, the results we report for the polygonal Lorentz model
were obtained with a square central obstacle, that is, a scatterer
whose opposite sides are parallel.  This alone does not, however,
imply that the system exhibits anomalous diffusion, since it is not
possible for a particle to propagate by colliding only with (copies
of) this square, i.e.\ the scatterers making up the square are not `accessible' from
one another. 

Any particle must in fact also collide with the
top and bottom scattering walls. Only if these walls are also
accessibly parallel as defined above (i.e.\ have parallel partners in
the unit cell) do we find anomalous diffusion.  Furthermore, if we
alter the central obstacle shape so that it no longer has parallel
sides, then we still find anomalous diffusion provided that the scattering
walls still have parallel partners which are accessible from one cell to another.

\begin{figure}
\scalefig{0.9}{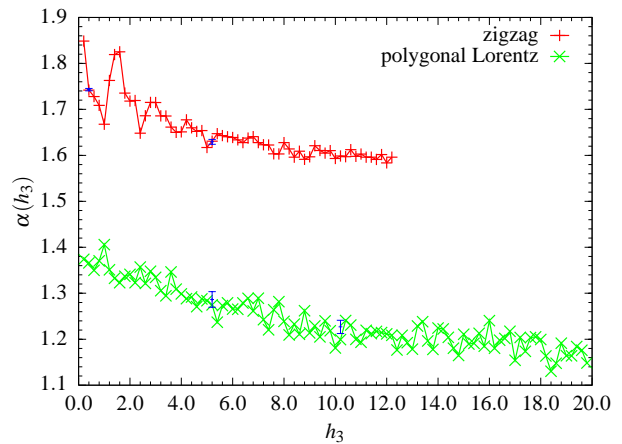} 
\caption{\label{fig:exponents} (Color online) Diffusion exponent
$\alpha$ as a function of $h_3$ in the parallel zigzag model with $h_1
= 0.1$ and the parallel polygonal Lorentz model with $h_1 = 0.1$ and
$w = 0.15$, calculated using $10^5$ initial conditions evolved until
the presumed asymptotic regime.  (Note that at $h_3 = 1 = d$ there is
a rational angle in both models.)  Representative error bars show the
standard deviation of the exponent over $100$ independent simulations.
}
\end{figure}

\bfigref{fig:exponents} shows the behavior of the anomalous diffusion
exponent $\alpha$ as a function of $h_3$ in the parallel zigzag and
polygonal Lorentz models with finite horizon.  The representative
error bars shown give the standard deviation of the exponent over
$100$ independent simulations for the same value of $h_3$, and
indicate that the structure visible in the curve for the zigzag model
is not an artifact.

\begin{figure}
\scalefig{0.9}{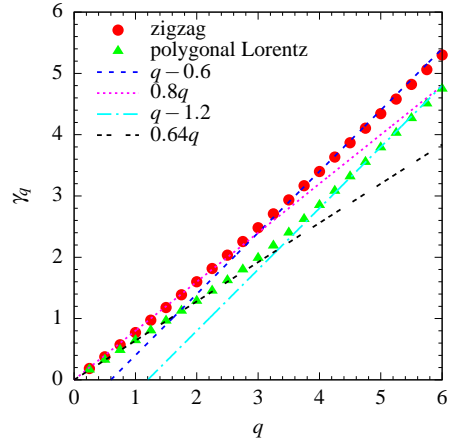} 
\caption{\label{fig:anom-moments} (Color online) Growth exponent
$\gamma_q$ of moments for the parallel polygonal Lorentz and zigzag
models with $h_1=0.51$, $h_2=h_3=10.0$ and $w=0.21$.  }
\end{figure}

\bfigref{fig:anom-moments} shows the growth exponent $\gamma_q$ as a
function of $q$ in two models with anomalous diffusion.  Again we find
a crossover between two linear regimes, with $\gamma_q$ of the form
\cite{CastiglioneStrongAnomDiffn}
\begin{equation}
\gamma_q = \begin{cases}
		\nu_1 q, & \text{for } q < q\crit\\
		q - \nu_2, & \text{for } q > q\crit.
           \end{cases}
\end{equation}

\subsection{Time evolution of densities}

A feature typical of normal diffusive systems is that the probability
distribution of displacements (or equivalently of positions) converges
in the long-time limit, when rescaled by $\sqrt{t}$, to to a normal
distribution \cite{Sanders2005}.  We find this behavior numerically in
both of the classes of polygonal billiard channels studied here, in
the cases for which the mean squared displacement grows linearly, and
we conjecture that this holds in general, providing a stronger sense
in which polygonal billiards can be considered diffusive.

\begin{figure}
\scalefig{0.9}{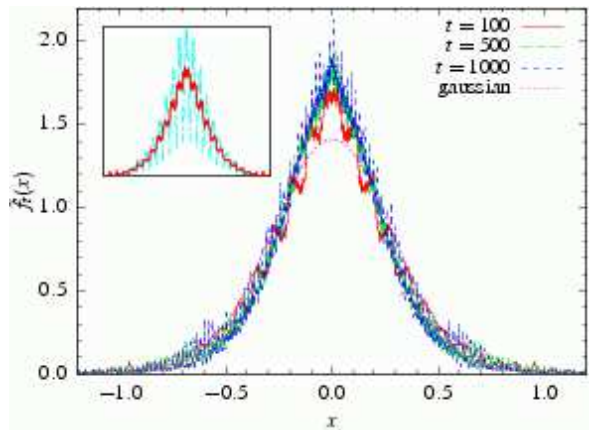}
\caption{\label{fig:anom-rescaled} (Color online) Rescaled demodulated
densities in the polygonal Lorentz model, compared to a Gaussian with
the corresponding variance. The inset shows the original and
demodulated (heavy line) densities for $t=100$. The model parameters
are $h_1=0.1$, $h_2=h_3=0.45$, $w=0.2$, $d=1$.}
\end{figure}

For anomalous diffusion with $\mean{\Dx^2}_t \sim t^\alpha$ ($\alpha >
1)$, we no longer expect the central limit theorem to hold.  The inset
of \figref{fig:anom-rescaled} shows the probability density
$\rho_t(x)$ of positions $x(t)$ in a parallel polygonal Lorentz model.
It exhibits a striking fine structure, similar to that found in
normally diffusive Lorentz gases and polygonal billiards in
\cite{Sanders2005}, where it was shown that the principal contribution
to this fine structure arises from a geometrical effect, namely that
the amount of vertical space available for particles to occupy varies
along the channel.  Assuming that the dynamics has good mixing
properties, it was shown that defining the \defn{demodulated} density
$f_t(x)$ by
\begin{equation}
f_t(x) \defeq \rho_t(x) / h(x)
\end{equation}
eliminates much of the fine structure.  Here, $h(x)$ is the height of
the available space in the billiard channel at horizontal coordinate
$x$, normalised by the area of the unit cell so that the integral
$\int_{-d}^d h(x) \rd x$ over the unit cell is $1$.  This demodulated
density is also shown in the inset of \figref{fig:anom-rescaled},
although the demodulation is not as effective as in the normally
diffusive cases studied in \cite{Sanders2005}, due to the weaker
mixing properties of the current model.

Rescaling the demodulated position density $f_t(x)$ at time $t$ by 
\begin{equation}
\tilde{f}_t(x) \defeq t^{\alpha/2} \, f_t(x \, t^{\alpha/2})
\end{equation}
gives a bounded variance as $t \to \infty$ \cite{ProsenAnomDiffn}.
The main part of \bfigref{fig:anom-rescaled} shows demodulated
densities rescaled in this way for different times, compared to a
Gaussian with the corresponding variance.  The rescaled densities
converge at long times to a non-Gaussian limiting shape, as has
previously been found in other systems with anomalous diffusion
\cite{MetzlerKlafter}.

\subsection{Mechanism for anomalous diffusion}

\subsubsection{Continuous-time random walk approach}

\begin{figure}
\scalefig{0.9}{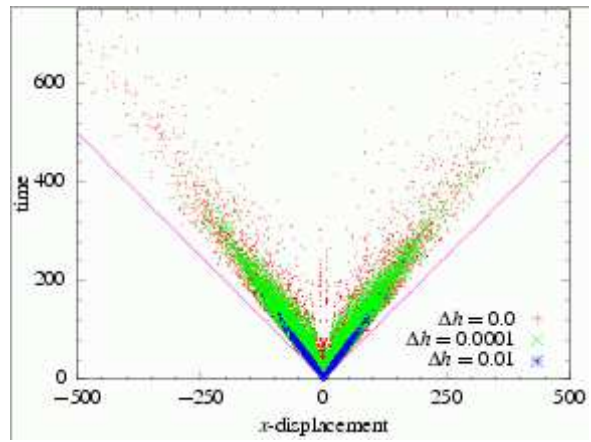} 
\caption{\label{fig:zigzag-laminar-joint} (Color online)
Scatterplot representing the
joint density $\psi(x,t)$ of laminar stretches in the zigzag model
with $h_1=0.1$, $h_3=0.3$ and $h_2 = h_3 + \Delta h$. The straight
lines with slope $1$ correspond to the maximum possible speed.
}
\end{figure}

For the parallel zigzag model, \bfigref{fig:periodic-orbits}
demonstrates the presence of long \defn{laminar stretches}, i.e.\
periods of motion during which a particle's velocity component
parallel to the channel wall does not change sign.  Typical
trajectories exhibit such behavior, sometimes over very long periods
of time.  \bfigref{fig:zigzag-laminar-joint} shows a scatterplot
representing the joint distribution $\psi(x,t)$ of laminar stretches
of duration $t$ with horizontal displacement $x$ along the
channel. (To compute these data, the channel was divided into sections
lying between adjacent maxima and minima in order to determine the
wall direction).

Thus, we may attempt to describe the dynamics of this system as a
\defn{continuous-time random walk} (CTRW), whose steps are the laminar
stretches: see e.g.\ \cite{WeissRandomWalk, ZumofenKlafter}.  To this
end, we note that \bfigref{fig:zigzag-laminar-joint} shows that the
distribution $\psi(x,t)$ is concentrated on diagonal lines emanating
from the origin, which leads us to try an approximation of the form
$\psi(x,t) = \textstyle \frac{1}{2} \delta(\modulus{x}-vt) \, \psi(t)$ for some
speed $v$.  This version of the CTRW was termed the \emph{velocity
model} in \cite{ZumofenKlafter}, and describes motion at a constant
velocity for a time $t$ in the direction $x$; after each stretch the
direction is randomised and a new step is taken.  With this form for
$\psi(x,t)$, the long-time growth of the mean squared displacement
depends on the asymptotic decay rate of the marginal distribution
$\psi(t)$ of step times \cite{ZumofenKlafter}: if $\psi(t) \sim
t^{-1-\nu}$ as $t \to \infty$, then the variance $\sigma^2(t)$ behaves
like $\sigma^2(t) \sim t^{3-\nu}$ when $1<\nu<2$, while we have normal
diffusion $\sigma^2(t) \sim t$ for $\nu>2$.

\begin{figure}
\scalefig{0.9}{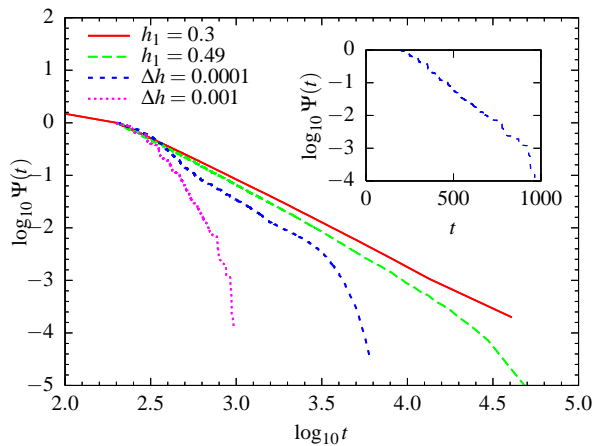} 
\caption{\label{fig:zigzag-laminar-t-distn} (Color online) Tail of
$\Psi(t)$ for the parallel zigzag model.  The top curve is for
$h_1=0.1$ and $h_3 = 0.3$.  The lower three curves are for $h_1 =
0.49$, $h_3 = 5.1$ and $\Delta h = 0$, $0.0001$ and $0.001$ (top to
bottom).  The inset is a semilogarithmic plot of the $\Delta h =
0.001$ case, showing exponential decay.}
\end{figure}

\bfigref{fig:zigzag-laminar-t-distn} shows the tail region of $\Psi(t)
\defeq 1 - \int_0^t \psi(\tau) \rd \tau$.  From the top curve we find
a long-time power-law decay $\Psi(t) \sim t^{-1.58}$ for a particular
case having parallel scatterers. For close-to-parallel configurations,
the tail of the distribution follows that for the parallel case, but
with an exponential cut-off, as shown in the inset of
\figref{fig:zigzag-laminar-t-distn}. According to the CTRW model, this
cut-off gives rise to normal diffusion in the asymptotic long-time
regime. This change of behavior is in qualitative agreement with our
observations. Unfortunately, the value of the tail exponent for the
parallel case corresponds to $\nu=1.58$ and hence to a mean squared
displacement $\sigma^2(t) \sim t^{3-1.58}=t^{1.42}$, which does not
agree with the numerically observed growth rate $\sigma^2(t) \sim
t^{1.81}$.

Although the above results rest on a rather gross approximation to the
joint distribution $\psi(x,t)$, we doubt that a more general CTRW
model, incorporating information on the complete $\psi(x,t)$, would
substantially improve the agreement. The reason is that correlations
between consecutive laminar trajectories have an important effect not
accounted for by CTRW models.  Indeed, for a given trajectory, we find
that the lengths of consecutive long laminar stretches are often
nearly equal, indicating that the trajectory repeats very closely its
previous behavior many times.

\subsubsection{Propagating periodic orbits}

\begin{figure}
\scalefig{0.9}{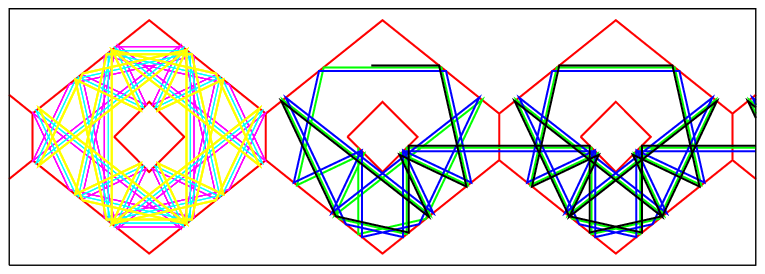}
\scalefig{0.9}{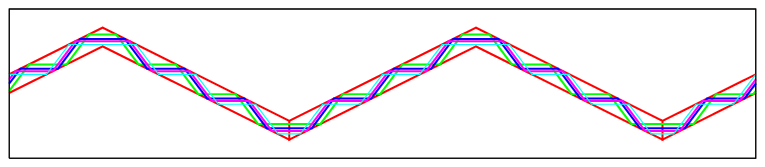}
\caption{\label{fig:periodic-orbits} (Color online) Families of
trapped and propagating periodic orbits in parallel systems.}
\end{figure}

On the other hand, our results indicate that it is the presence of
accessible parallel scatterers that gives rise to anomalous
diffusion. We believe that \emph{propagating periodic orbits}, i.e.\
orbits which repeat themselves with a spatial displacement, may
provide a mechanism for this behavior.  Such orbits are more prevalent
when there are parallel scatterers, since in this case it is much
easier for a particle to regain its original angle of propagation
after a given number of bounces, and, then, possibly repeat its
previous motion. In addition, periodic orbits in polygonal billiards
occur in families \cite{GalperinPeriodicOrbitsBilliards}; several such
families are shown in \figref{fig:periodic-orbits}.

Furthermore, trajectories with initial conditions which are close in phase
space to those of a propagating periodic orbit will shadow it, and
hence propagate, for a long time, with a longer shadowing time for
closer initial conditions.  Anomalous diffusion should then result from
a balance between the ballistic propagation of the periodic orbits
themselves, the long-lasting ballistic motion of shadowing orbits, and
the diffusive motion of other trajectories.  

We have, however, not yet been able to account analytically for
anomalous diffusion in this way. For instance, the mere existence of
propagating orbits is not enough to give rise to the kind of anomalous
behavior we observe. Indeed, families of propagating orbits exist in
the corridors of the infinite-horizon periodic Lorentz gas, but these
give rise only to marginally anomalous diffusion $\msd \sim t \ln t$
\cite{Bleher} (see \secref{sec:normal-diffn}). Thus, although the set
of propagating periodic orbits must have measure zero in phase space,
since otherwise their ballistic nature would result in overall
ballistic transport with $\msd \sim t^2$, they must be plentiful in
some sense, in order to give rise to anomalous transport. For example,
such orbits could be dense at least in some parts of phase space.

\section{Crossover from normal to anomalous diffusion}
\label{sec:crossover-normal-anom}

Since anomalous diffusion occurs only for special geometrical
configurations, it is of interest to study the 
crossover from normal to anomalous diffusion as such
configurations are approached. We begin by considering the polygonal
Lorentz channel, in which we fix all geometrical parameters except for
$h_2$, which we vary as $h_2 = h_3 + \Delta h$.

\begin{figure}[t]
\scalefig{0.9}{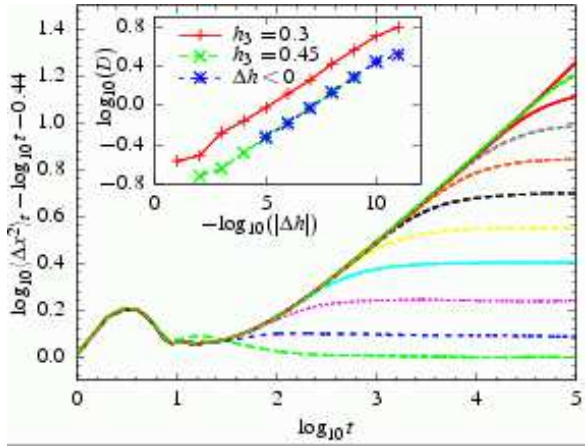}
 \caption{\label{fig:crossover} (Color online) Double logarithmic plot
 of $\msd_t$ as a function of time for the polygonal Lorentz model
 with $h_1=0.1$, $h_3=0.45$, $w=0.2$ and $h_2 = h_3 + \Delta h$.
 Deviations from an asymptotic linear fit to the lowest curve are
 shown for $\Delta h=0$ (uppermost) and $\Delta h = 10^{-m}$,
 $m=2,\ldots,11$ (bottom to top).  The inset shows a double
 logarithmic plot of the diffusion coefficient $D$ as a function of
 $\Delta h$ for the polygonal Lorentz model with $h_1 = 0.1$ for
 $h_3=0.3, 0.45$, and for $h_3=0.45$ with $\Delta h < 0$.  }
\end{figure}

\bfigref{fig:crossover} shows a double logarithmic plot of the mean
squared displacement in a polygonal Lorentz model for values of
$\Delta h$ tending to $0$, i.e.\ approaching the parallel
configuration; a linear fit to the bottom curve has been subtracted
from each curve for clarity.  As $\Delta h$ tends to $0$, the curve
for the parallel configuration ($\Delta h=0$) is followed for
progressively longer times before a crossover occurs to asymptotic
diffusive behaviour, which appears on the figure as a zero slope.
Defining the crossover time $\tc(\Delta h)$ as the intersection of the
initial anomalous growth with a straight line fit to the asymptotic
normal growth, we find that $\tc$ scales like $\tc \sim (\Delta
h)^{-0.37}$ as $\Delta h \to 0$.

The inset of \figref{fig:crossover} shows the diffusion coefficient
$D$ as a function of $\Delta h$, obtained from the asymptotic slope of
the mean squared displacement.  Although \figref{fig:crossover} shows
that the asymptotic linear regime has not been reached for the
smallest values of $\Delta h$, thereby underestimating the diffusion
coefficient, we obtain a straight line on a double logarithmic plot of
$D$ versus $\Delta h$, giving the power-law behavior
\begin{equation}\label{}
D(\Delta h) \sim  
\modulus{\Delta h}^{-0.15} \quad \text{as
} \modulus{\Delta h} \to 0,
\end{equation}
with the same exponent for both positive and negative values of
$\Delta h$ and for two (close) values of $h_3$. For this model, the
diffusion exponent in the parallel ($\Delta h=0$) case is $\alpha
\simeq 1.40$.
 
A similar crossover to anomalous diffusion occurs near to the parallel
configuration in the zigzag model (not shown), and we conjecture that
such crossover behavior is generally found in polygonal billiards near
to configurations exhibiting anomalous diffusion.  The rate of growth
of the diffusion coefficient for $\Delta h \neq 0$ depends, however,
on the diffusion exponent for $\Delta h = 0$: for the zigzag model
with $h_1 = 0.1$ and $h_3 = 0.3$, we find $D \sim \modulus{\Delta
h}^{-0.32}$, with a diffusion exponent $\alpha \simeq 1.81$ for the
parallel case; both of these exponents differ from the results in the
polygonal Lorentz model.

The growth of the diffusion coefficient close to a parallel case thus
depends on the anomalous diffusion exponent found in that parallel
case. Indeed, the relation between the exponents characterizing the
crossover from normal to anomalous transport can be obtained from the
following simple scaling argument. Define the crossover time exponent
$\nu$ by $\tc\sim |\Delta h|^{-\nu}$.  We have
\begin{equation}
\langle x^2(\Delta h)\rangle_t \sim \begin{cases}
		D(\Delta h)t, & \text{for } t > \tc \\
		t^\alpha, & \text{for } t < \tc,
           \end{cases}
\end{equation}
so that continuity at $\tc$ gives $D(\Delta h)\sim \tc^{\alpha-1}\sim
|\Delta h|^{\nu(1-\alpha)}$,  which is in good quantitative
agreement with our results. 

\begin{figure}
\scalefig{0.9}{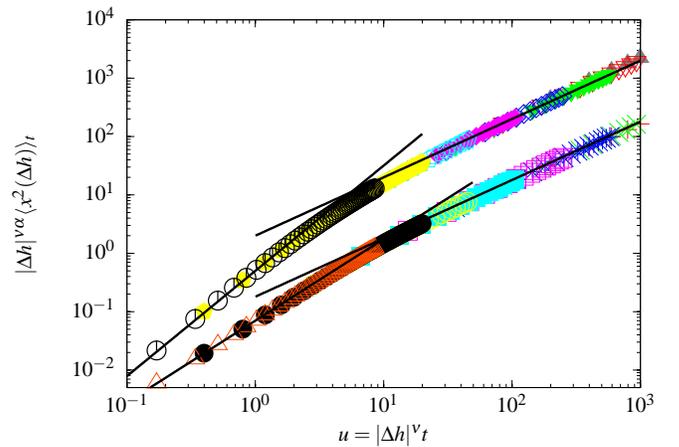} 
\caption{\label{fig:collapse} (Color online) Data collapse of $|\Delta
h|^{\nu\alpha}\langle x^2(\Delta h)\rangle_t$ as a function of
$u=|\Delta h|^{\nu} t$ for the zigzag model with $h_1=0.1$, $h_3=0.3$,
which has $\alpha \simeq 1.81$ (upper curve) and the polygonal Lorentz
model with $h_1 = 0.1$, $h3=0.45$, with $\alpha = 1.40$.  In both
models the value $\nu=0.37$ was used, and curves are shown for $\Delta
h=10^{-m}$, $m=4,\ldots,11$.  The solid lines show the extreme
behaviors of the scaling function \eqref{eq:crossover-function}.}
\end{figure}

Furthermore, this argument implies that plots of $|\Delta
h|^{\nu\alpha}\langle x^2(\Delta h)\rangle_t$ as a function of $u
\defeq |\Delta h|^{\nu} t$ should collapse onto a scaling function of
the form
\begin{equation}\label{eq:crossover-function}
\phi(u)\sim \begin{cases} u, & \text{for } u \gg 1\\
                      u^{\alpha},  & \text{for } u \ll 1.
           \end{cases}
\end{equation}
The data collapse for both the zigzag and the polygonal model near a
configuration of parallel scatterers is shown in
\figref{fig:collapse}. Interestingly, while the transport exponent
$\alpha$ is sensitive to the geometrical details of the models, the
best collapse was obtained for the same value of $\nu=0.37$ in both
cases.

Finally, it should be noted that the above calculations all refer to
irrational angles; the situation when the value of $\Delta h$ renders
an angle rational is unclear.

\section{Discussion and Conclusions}

We have demonstrated that the type of diffusive behavior exhibited by
periodic polygonal billiard channels depends on geometrical features
of the unit cell.  Based on our results, we conjecture that sufficient
conditions for a periodic polygonal billiard channel to exhibit normal
diffusion are the following: (i) all vertex angles are irrationally
related to $\pi$; (ii) the billiard has a finite horizon; and (iii)
there are no accessible parallel scatterers in the unit cell.

When there is an infinite horizon, diffusive transport is replaced by
marginal anomalous diffusion with $\msd \sim t \ln t$, while if there
are accessible parallel scatterers, then anomalous diffusion with
$\msd \sim t^\alpha$, $\alpha > 1$ is observed.

In the zigzag model we also have evidence of anomalous diffusion
(e.g.\ with exponent $1.16$ for one particular model) with irrational
angles chosen such that one angle is twice the other, i.e.\ with a
rational relation between them, in disagreement with the result found
in \cite{RondoniZigzagPreprint} (where a rather small number of
initial conditions was used).  For other angle ratios we do not have
conclusive data, but the exponents are seemingly close to $1$, while
for the polygonal Lorentz model we find normal diffusion in similar
situations.  For the zigzag model we must thus also exclude the
possibility of rationally-related angles from the above conditions for
normal diffusion.

We remark that in \cite{AlonsoPolyII}, propagating periodic orbits were
conjectured to be related to anomalous diffusion in a system with one
rational angle, but no reason was given for their existence.  Actually, 
it is possible to include that system into our picture by unfolding
the rational angle, which gives rise to an equivalent system with
parallel scatterers.  

We believe that the explanation of anomalous diffusion should be found
in terms of such propagating periodic orbits, which are much more
prevalent in the presence of parallel scatterers.  We further showed
that there is a crossover from normal to anomalous diffusion as a
parallel configuration is approached, with the diffusion coefficient
having a power-law divergence.  We hope to achieve a quantitative
description of both of these points in the future.

\acknowledgments

This work was initiated as part of the first author's PhD thesis
\cite{SandersThesis}; he thanks his supervisor Robert MacKay for
valuable comments, and UNAM for a postdoctoral research award.  The
Centre for Scientific Computing at the University of Warwick provided
computing facilities for some of the calculations. Support through
grant IN-100803 DGAPA-UNAM is also acknowledged.

\def\cprime{$'$} \def\cprime{$'$}

\end{document}